\documentclass[aps,prl,showpacs,preprint,superscriptaddress]{revtex4}

\usepackage{graphicx}
\usepackage{dcolumn}
\usepackage{bm}
\usepackage{units}
\usepackage{color}
\begin{document}

\title{Itinerant Nature of Atom-Magnetization Excitation by Tunneling Electrons}

\author{A. A. Khajetoorians}
\affiliation{Institute of Applied Physics, Hamburg University, Jungiusstrasse 11, D-20355 Hamburg, Germany}
\author{S. Lounis}
\affiliation{Department of Physics and Astronomy, University of California Irvine, California, 92697 USA}
\author{B. Chilian}
\affiliation{Institute of Applied Physics, Hamburg University, Jungiusstrasse 11, D-20355 Hamburg, Germany}
\author{A. T. Costa}
\affiliation{Department of Physics and Astronomy, University of California Irvine, California, 92697 USA}
\affiliation{Instituto de F\'isica, Universidade Fedeal Fluminense, 24210-340 Niter\'oi, RJ, Brazil}
\author{L. Zhou}
\affiliation{Institute of Applied Physics, Hamburg University, Jungiusstrasse 11, D-20355 Hamburg, Germany}
\author{D. L. Mills}
\affiliation{Department of Physics and Astronomy, University of California Irvine, California, 92697 USA}
\author{J. Wiebe}
\email{jwiebe@physnet.uni-hamburg.de}
\affiliation{Institute of Applied Physics, Hamburg University, Jungiusstrasse 11, D-20355 Hamburg, Germany}
\author{R. Wiesendanger}
\affiliation{Institute of Applied Physics, Hamburg University, Jungiusstrasse 11, D-20355 Hamburg, Germany}

\date{\today}

\begin{abstract}
	We have performed single-atom magnetization curve (SAMC) measurements
	and inelastic scanning tunneling spectroscopy	(ISTS)
	on individual Fe atoms on a Cu(111) surface. The SAMCs show a broad distribution of magnetic
	moments with $\unit[3.5]{\mu_{\rm B}}$ being the mean value.
	ISTS reveals a magnetization excitation with a lifetime of
	$\unit[200]{fsec}$ which decreases by a factor of two
	upon application of a magnetic field of $\unit[12]{T}$.
	The experimental observations are quantitatively explained by
	the decay of the magnetization excitation into Stoner modes
	of the itinerant electron system as shown by newly developed theoretical modeling.
\end{abstract}

\maketitle

Magnetic atoms adsorbed on nonmagnetic substrates have been a topic of active study, both to provide insight into fundamental aspects of magnetism and as possible elements for future information
technology and spin-based computation schemes. Depending on the type of substrate, ranging from thin insulating layers~\cite{Heinrich2004,Hirjibehedin2007,Loth2010} over semiconductors~\cite{Khajetoorians2010}, to metals~\cite{Gambardella2003,Meier2008,Balashov2009,Zhou2010}, one expects increasing hybridization of the atom with
the substrate states. The itinerant nature of the substrate electrons play an increasingly pronounced role in the static and dynamic properties of the magnetic atom as one progresses through this sequence. Spin-polarized (SP-STS)~\cite{Meier2008} and inelastic (ISTS)~\cite{Heinrich2004} scanning tunneling spectroscopy or a combination of both~\cite{Loth2010,Khajetoorians2010} provide the only current means to probe magnetic properties and spin dynamics of isolated magnetic adsorbates on the atomic-scale.

For magnetic atoms weakly hybridized with a non-magnetic surface, a description of the atomic moment by a half-integer spin governed by anisotropy terms in a spin Hamiltonian (``isolated spin model'') is sufficient~\cite{Heinrich2004,Hirjibehedin2007,Loth2010,Khajetoorians2010, Lorente2009}.  Within this approximation, the role of the underlying host conduction electrons is neglected and the effect of the substrate is encompassed in the magnetic anisotropy.  However, it has been known for decades that the  magnetic moments of 3d impurities in a nonmagnetic 3d, 4d or 5d metal are influenced qualitatively by the itinerant conduction electrons of the host material. Accordingly, as theoretically predicted a long time ago, the magnetization dynamics of such systems is damped by decay into electron-hole pairs, namely Stoner excitations of the itinerant electron gas \cite{Mills1967}. One consequence of this decay is a substantial $g$-shift and energy-dependent linewidth of the magnetization excitation if detected via a local method such as STS~\cite{Mills1967,Muniz2003}. Therefore the application of the isolated spin model to magnetic atoms on metals, while descriptive, inadequately describes these types of effects. This view is reinforced by the large excitation linewidth observed in previous ISTS which cannot be described by an approach which treats the magnetic atom as an isolated entity governed solely by magnetic anisotropy ~\cite{Balashov2009,schuh:09E156}. While the effects of Stoner excitations have been observed experimentally by spin-polarized electron-energy-loss spectroscopy of ferromagnetic surfaces~\cite{Hopster_PhysRevLett.53.695,Kirschner_PhysRevLett.55.973}, an experimental verification of their importance for the magnetization dynamics of individual impurities is thus far lacking.

In this letter, we reveal the itinerant nature of individual Fe atoms absorbed on a Cu(111) surface utilizing a combination of single-atom magnetization curve (SAMC) measurements \cite{Meier2008} as probed by SPSTS and tunneling-electron driven excitations (ISTS).
SAMCs reveal the magnetic moment of individual Fe atoms which is $\approx 3.5 \mu_{B}$. Complementary to this technique, ISTS reveals an anisotropy gap as well as a large linewidth of the magnetization excitation which increases linearly with magnetic field. In addition, we present the first theoretical studies which incorporate spin-orbit coupling in the response of an isolated local moment hybridized strongly with the host conduction electrons. The calculations provide a quantitative account of the anisotropy gap, and the linear variation with magnetic field of both linewidth and peak energy in the magnetization excitation spectrum. This confirms that the excitation lifetime is limited by decay into Stoner modes with a density of states linearly increasing in energy as predicted earlier~\cite{Mills1967,Muniz2003}.

\begin{figure}[t]
\centerline{\includegraphics[width = \columnwidth]{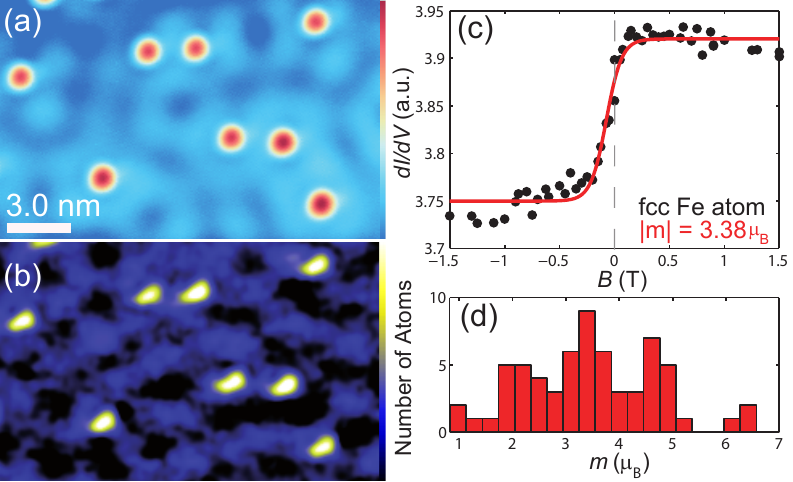}}
\caption{(color online). (a) Topograph of a distribution of Fe atoms on Cu(111). The colorbar spans $\unit[90]{pm}$. (b) Spatially-resolved magnetic asymmetry map between saturation field values $B=\unit[\pm 0.4]{T}$. ($I_{\rm stab}=\unit[0.6]{nA}$, $V_{\rm stab}=\unit[-10]{mV}$, $V_{\rm mod}=\unit[5]{mV}$~(rms)).  Color bar from 0 {a.u.} to 0.025 {a.u.} (c) Dots: SAMC extracted from the $B$-dependent ${\rm d}I/{\rm d}V$ maps by averaging over the signal from an individual atom. Solid line: fit to the data using the continuum model. (d) Histogram of effective magnetic moments extracted from the fit to the SAMCs of Fe atoms with minimum distance $\unit[2]{nm}$ to other atoms or Co islands.\label{fig:mag}}
\end{figure}

All experiments were performed on a home-built UHV STM with a base temperature of $T=\unit[0.3]{K}$ capable of applying a magnetic field $B$ perpendicular to the surface~\cite{Wiebe2004a}.
Non-spin sensitive W tips were used for ISTS and Cr-coated W tips were used for SAMC measurements~\cite{Kubetzka2002,Meier2008}. The Cu(111) surface was cleaned by repeated Ar$^+$ sputtering and annealing cycles and Co islands were deposited at room temperature with a nominal coverage of $\unit[0.5]{ML}$~\cite{delaFiguera1993}. Such islands have single out-of-plane magnetized domains and were utilized to confirm the out-of-plane sensitivity of each Cr-tip ~\cite{Pietzsch2004}. Subsequently, Fe was deposited onto the cold surface at $T < \unit[6]{K}$~\cite{Crommiecorral}.
This results in a statistical distribution of single Fe atoms with a nominal coverage of a few thousandths of a monolayer (Fig.~\ref{fig:mag}(a)). The atoms were observed to occupy only one adsorption site~\cite{Stroscio2004}. STM topographs were recorded in constant-current mode at a stabilization current $I_{\rm stab}$ with a bias voltage $V_{\rm stab}$ applied to the sample. SAMCs are recorded as described in Ref.~\cite{Meier2008} and in the supplementary material~\cite{supplement}. For ISTS, ${\rm d}I/{\rm d}V(V)$ is recorded in open feedback mode via a lock-in technique with a small modulation voltages $V_{\rm mod}$ added to the bias ($f = \unit[4.1]{kHz}$).

In order to determine the magnetic moment of the individual Fe atoms, we measured the SAMCs of several single atoms (Fig.~\ref{fig:mag}(a)) using an out-of plane magnetized tip~\cite{Meier2008}. ${\rm d}I/{\rm d}V$ maps acquired at various $B$ of the same area reveal magnetic contrast on top of the Fe atoms~\cite{supplement}. From such maps, the spatially-resolved magnetic asymmetry is calculated by  $({\rm d}I/{\rm d}V (B_{\uparrow}) - {\rm d}I/{\rm d}V (B_{\downarrow})) / ({\rm d}I/{\rm d}V (B_{\uparrow}) + {\rm d}I/{\rm d}V (B_{\downarrow}))$ as given in Ref.~\cite{Bode2002}. Fig.~\ref{fig:mag}(b) illustrates a magnetic asymmetry map evaluated at opposite saturation fields ($B_{\uparrow} = -0.4$ T, $B_{\downarrow} = +0.4$ T).  Each SAMC (e.g. Fig.~\ref{fig:mag}(c)) is acquired by extracting ${\rm d}I/{\rm d}V$, for a given atom, during a $B$ sweep (each $B$ value corresponds to data extracted from one ${\rm d}I/{\rm d}V$ map). Similar curves have been measured for about 60 different atoms at different locations with different spin-polarized microtips where each atom probed had a minimum distance of 2 nm from any other magnetic structure. All reveal paramagnetic behavior with a saturation at $B_{\rm sat}\approx0.2$~T.

Because of the hybridization of the atom with the substrate, a quasiclassical continuum model~\cite{Gambardella2003,Meier2008} is appropriate to describe the measured SAMCs. Fits utilizing various models produce a similar distribution of moments as a result of the rather significant out-of-plane anisotropy~\cite{Lazarovits_PhysRevB.68.024433}. This model properly reproduces the measured curve if we assume an effective magnetic moment $m\approx\unit[3.5]{\mu_{B}}$ with a magnetic anisotropy of $\approx 1$ meV (Fig.~\ref{fig:mag}(c)). The histogram of fitted $m$ for all measured atoms is shown in Fig.~\ref{fig:mag}(d). The average value is in good agreement with the DFT calculated total magnetic moment which considers both the spin and orbital moments of the Fe and the neighboring Cu atoms~\cite{Lazarovits_PhysRevB.68.024433,Lounis2006}. However, there is a broad distribution of the effective $m$. We speculate on two substrate-mediated effects: (i) there is a spatially varying mean field due to the inter-atomic \emph{long range} RKKY interaction which mimics a different effective $m$ for each atom~\cite{Meier2008,Zhou2010}. (ii) The varying substrate density of states resulting from surface-state electron scattering might additionally change the value of $m$. While the detailed distribution of $m$ changes when we use different minimum separation cutoffs, the maximum is always centered around $\unit[3.5]{\mu_{B}}$ independent of this choice.

\begin{figure}[t]
\centerline{\includegraphics[width = \columnwidth]{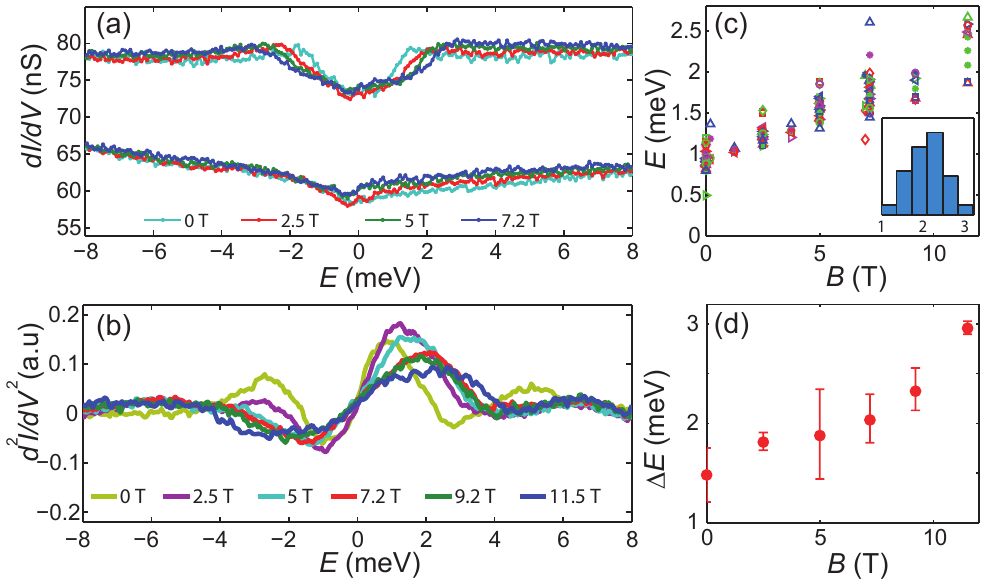}}
\caption{(color online). (a) ISTS at different $B$ as indicated ($I_{\rm stab}=\unit[1]{nA}$, $V_{\rm stab}=\unit[-10]{mV}$, $V_{\rm mod}=\unit[0.1]{mV}$~(rms)). Top spectra were taken on an Fe atom. Bottom spectra were taken with the same tip on Cu (vertically shifted by $\unit[-14]{nS}$). (b) Numerical differentiation of spectra taken on another Fe atom using a different tip ($I_{\rm stab}=\unit[2]{nA}$, $V_{\rm stab}=\unit[-10]{mV}$, $V_{\rm mod}=\unit[0.05]{mV}$~(rms)). (c) Energy and (d) averaged FWHM (high resolution data; error corresponds to the maximum and minimum value) of the excitation as a function of $B$ extracted from gaussian fits to spectra (b). The inset in (c) shows a histogram of the measured $g$-factors determined from the slope of (c) divided by $\mu_{\rm B}$. The values are extracted from spectra measured on 78 (c) and 16 (d) different atoms.\label{fig:spectra}}
\end{figure}

Complementary to SAMC measurements, ISTS can reveal information about the dynamical magnetic properties of the Fe atoms. Fig.~\ref{fig:spectra}(a) shows ISTS curves taken on an isolated Fe atom (top) and on the Cu(111) substrate (bottom) using the same tip atom as the magnetic field $B$ is varied. While the substrate spectra are relatively flat and show no clear trend in a field, the atom spectra always show two steps symmetric with respect to the Fermi level $E_{\rm F}$ ($V=\unit[0]{mV}$). These shift linearly towards higher voltage with increasing $B$ and are accompanied by a simultaneous increase in linewidth. Numerical differentiation of the spectra illustrates both effects (Fig.~\ref{fig:spectra}(b)). Slight asymmetries in the lineshape for positive and negative voltage originate from small features in the tip or substrate density of states (Fig.~\ref{fig:spectra}(a) bottom). Similar spectra using different tips were recorded on tens of different atoms with high energy resolution ($\approx 150 \mu$eV). All atoms show the same behavior, although with small quantitative differences in energy and linewidth.

The peak energy $E$ and full width half maximum (FWHM) linewidth $\Delta E$ of the step was extracted from Fig.~\ref{fig:spectra}(b) utilizing gaussian functions and plotted as a function of $B$ in Fig.~\ref{fig:spectra}(c) and (d). As shown in the histogram in Fig.~\ref{fig:spectra}(c) (inset), the peak energy has an average slope of $\unit[2.1]{\mu_{\rm B}}$ (with a min/max of $\unit[2.1] \pm\unit[1]{\mu_{\rm B}}$). This proves that the observed steps in ${\rm d}I/{\rm d}V$ are due to magnetic excitations of each Fe atom induced by tunneling electrons~\cite{Heinrich2004,Hirjibehedin2007} resulting in a mean $g$-factor of $2.1$ for this system. The height of the steps reveal that only about $\approx 5$\% of all tunneling electrons induce such excitations ~\cite{Lorente2009}. The zero field excitation energy in the ISTS curves indicates that each Fe atom is subject to a substrate-induced magnetic anisotropy~\cite{Heinrich2004,Hirjibehedin2007} of about $\unit[1]{meV}$, which is four times smaller than the DFT predicted uniaxial out-of-plane anisotropy~\cite{Lazarovits_PhysRevB.68.024433}. This measured anisotropy value was in turn utilized in the SAMC fitting model above. In passing we remark that, since the energy of the tunneling electrons used for the SAMCs is larger than the excitation threshold, 5\% of the tunneling electrons excite the magnetization during recording the SAMC, leading to a time of $\unit[3]{nsec}$ between consecutive excitation events ($I_{\rm stab}=\unit[0.6]{nA}$). This time is six orders of magnitude shorter than the time resolution of the experiment and more than four orders of magnitude longer than the lifetime of the excitation, justifying the assumption of thermal equilibrium within the model used for the SAMC curve fitting.


Most notably, the excitation steps are very broad, with a width of $\Delta E = \unit[1.5]{meV}$ that increases linearly with $B$ as shown in Fig.~\ref{fig:spectra}(d).
Since artificial experimental broadening due to $V_{\rm mod}$ and a nonzero temperature~\cite{Stipe1998} is negligible in our experiment, the lifetime of the excitation is calculated
to be remarkably short $\tau=\hbar/(2\Delta E)=\unit[200]{fsec}$ ($B=\unit[0]{T}$). Comparably short lifetimes of magnetic excitations of single atoms have been reported for Co and Fe adsorbates on Pt(111) and ascribed to the interaction between the atom states and the substrate electrons~\cite{Balashov2009,schuh:09E156}. In these papers, the magnetic anisotropy and the relaxation mechanism were both described within an isolated spin model, neglecting itinerant effects, which is misleading in the case of strong hybridization. In contrast, both experiments reported here demonstrate the itinerant nature of the Fe/Cu(111) system, as highlighted by the energy-dependent linewidth of the magnetic excitation and by the non-half integer total angular momentum $J = m/(g\mu_\mathrm{B})=1.75$. Thus, a description of the Fe atom within an isolated spin model is not suitable.

We turn next to a description of our theoretical studies, which incorporate the itinerant effects evident in the data. In our description of the 
magnetization excitations, we address the frequency dependent local transverse spin susceptibility $\chi(\Omega)$. The imaginary part of this function provides us with the density of states of magnetization excitations. Thus, $\mathrm{Im}[\chi(\Omega)]$ can be directly compared with the data displayed in Fig.~\ref{fig:spectra}(b). Two approaches have been developed to compute $\chi$ for nanostructures:  an empirical tight-binding 
approach (ETB)~\cite{Muniz2003} and a method~\cite{Lounis2010,Lounis2010b} based on the time-dependent density functional 
theory within the Korringa-Kohn-Rostoker (KKR) scheme~\cite{KKR_2002}. 
While the former requires parameters, namely the various hopping integrals, the latter does not incorporate the challenging task of including the spin-orbit coupling which is essential to properly reproduce the experimental observations. Thus, we report here our
calculations obtained within the ETB approach taking full account of the spin-orbit interaction, which is a major advance from the theory presented in Ref.~\cite{Muniz2003}. Thus, the influence of magnetic anisotropy on the excitation
spectrum is incorporated fully in our analysis, and the spin-orbit contribution to the $g-$shift is present as well. We have extended the formal development described in Ref.~\cite{Costa2010} to the description of the local susceptibility of a moment-bearing atom.  
We also used the KKR-based scheme described in Ref.~\cite{Lounis2010} by mimicking the 
spin-orbit interaction with an external static magnetic field. Results for $\chi(\Omega)$ generated with such a method are in good accord with those obtained with the ETB scheme we present here.

Before discussing the theoretical magnetization excitations, we note that the two methods: KKR and ETB, with parameters fitted to reproduce the electronic structure obtained with KKR, 
lead to ground state moments of $\unit[3.2]{\mu_{B}}$ and $\unit[3.6]{\mu_{B}}$, respectively.

\begin{figure}
	\centerline{\includegraphics[clip,width=\columnwidth]{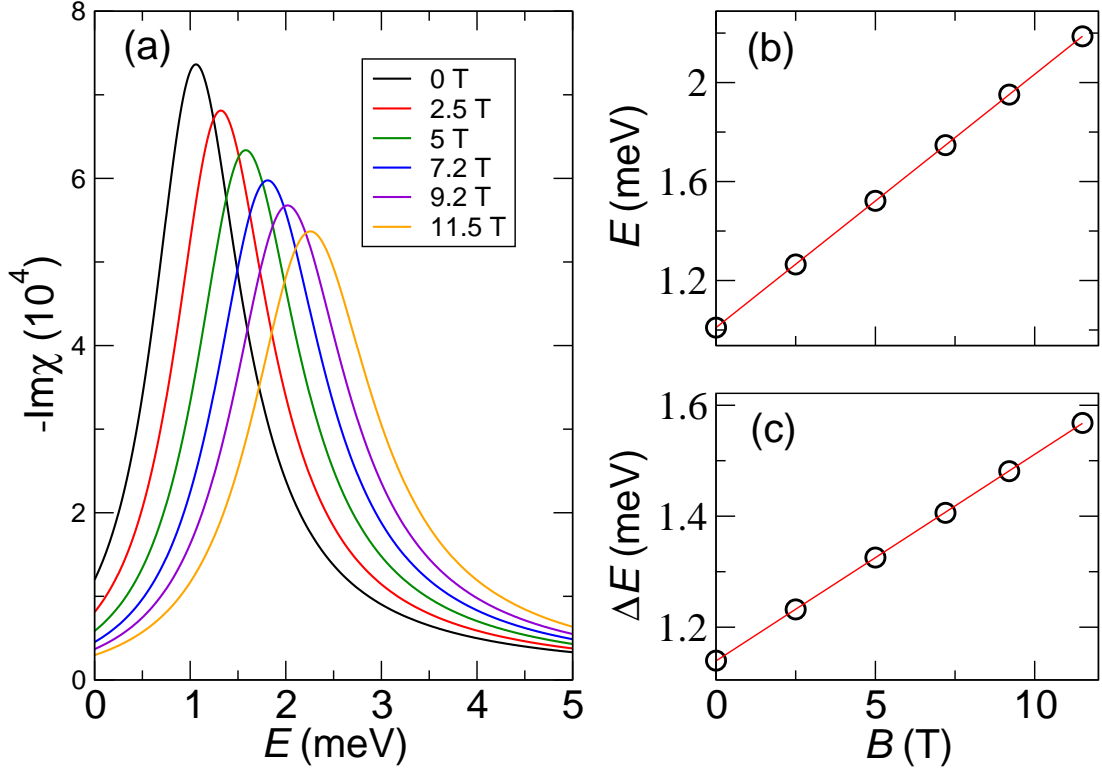}}
 	\caption{(color online) (a) Calculated density of
	magnetization excitations for an Fe atom on Cu(111) for various magnetic
	fields using the ETB approach. At zero field, spin-orbit induced anisotropy renders the excitation energy finite.
	(b) and (c) show magnetic field dependency of the peak energy and of the linewidth $\Delta E$ in the excitation spectrum (a).}
	\label{fig:calc}
\end{figure}

In Fig.~\ref{fig:calc}, we show the results of our theoretical studies, obtained with the ETB scheme, of the magnetization excitation spectrum of the Fe atom. In zero field, we see that the spin-orbit anisotropy renders the transition energy finite corresponding to a magnetic anisotropy of $\unit[1.1]{meV}$ which is very close to the experimental value (Fig.~\ref{fig:spectra}(c)). The spin-orbit coupling parameter used here is $\unit[34]{meV}$. This is 35\% lower than utilized in Ref.~\cite{Costa2010}, but compatible with the range of coupling constants found in the literature. Furthermore, use of a somewhat larger value of the spin-orbit coupling parameter, as discussed in Ref.~\cite{Costa2010}, leads to an anisotropy of $\unit[2.5]{meV}$ which is too large as compared to the experimental result.

The $B$ dependence of the peak energy in the excitation spectrum (Fig.~\ref{fig:calc}(b)) agrees nicely with the measured data. The theory gives a $g$-factor of 1.8, slightly smaller than the experimental mean. The absolute value of the linewidth and its increase with the magnetic field are qualitatively reproduced in the theory (Fig.~\ref{fig:calc}(c)), although with a smaller slope than in the measured data. A modest adjustment of the Fe/Cu hopping integrals would resolve this difference. The very large linewidths have their origin in the decay of the excitation into electron-hole pairs~\cite{Mills1967}. Consequently, the theory provides a very good account of the systematic features in the data.


In summary, with a combination of SAMC and ISTS measurements, we reveal the itinerant nature of the magnetization excitation of an Fe atom on the Cu surface. The signature of itinerant effects is manifested by a non-half integer magnetic moment and a large, magnetic field dependent, linewidth of its excitation. Calculations of the magnetic field dependent local dynamic susceptibility reveal that the excitation lifetime is dictated by a decay into Stoner modes of the itinerant electron system whose density of states is linearly increasing with energy and thus with magnetic field.  While Ref.~\cite{Mills1967} did not incorporate spin-orbit coupling and addressed moments in the bulk of a paramagnetic metal, the data reported here confirms the itinerant picture set forth in this paper. This picture will apply for other magnetic atoms on 3d, 4d or 5d metal surfaces.~\cite{Muniz2003,Lounis2010,Lounis2010b} It has been predicted that the coupling to Stoner modes is less efficient for (i) weak overlap of the d-minority states of the atom with $E_{\rm F}$ and (ii) weak overlap of the d-majority states with the d-bands of the substrate~\cite{Muniz2003,Lounis2010}. Such considerations consequently can serve to predict material combinations with long lifetimes of the magnetization excitation which is important for future spintronic applications.

We acknowledge funding from SFB668-A1 and GrK1286 of the DFG, from the ERC Advanced Grant ``FURORE", and from the Cluster of Excellence ``Nanospintronics". The research of D. L. M. and A. T. C. was supported by the U. S. Dept. of Energy, via grant DE-FG03-84ER-45083. S. L. thanks the Alexander von Humboldt Foundation for a Feodor Lynen Fellowship, and for partial support from the same DOE grant.

\end{document}